\documentclass[aps,prl,lengthcheck,superscriptaddress,reprint,showpacs,longbibliography,nofootinbib,amsmath,amssymb]{revtex4-1}
\usepackage[utf8]{inputenc}
\usepackage[T1]{fontenc}

\usepackage{amssymb}
\usepackage{graphicx}
\usepackage{grffile}
\usepackage{microtype}
\usepackage[usenames,dvipsnames]{xcolor}
\usepackage[colorlinks,linkcolor=blue,citecolor=blue,urlcolor=blue]{hyperref}

\def\equationautorefname~#1\null{Eq.~(#1)\null}

\begin{document}

\title{Many-molecule reaction triggered by a single photon in polaritonic chemistry}

\author{Javier Galego}
\affiliation{Departamento de F{\'\i}sica Te{\'o}rica de la Materia Condensada and Condensed Matter Physics Center (IFIMAC), Universidad Aut\'onoma de Madrid, E-28049 Madrid, Spain}
\author{Francisco~J.~Garcia-Vidal}
\email{fj.garcia@uam.es}
\affiliation{Departamento de F{\'\i}sica Te{\'o}rica de la Materia Condensada and Condensed Matter Physics Center (IFIMAC), Universidad Aut\'onoma de Madrid, E-28049 Madrid, Spain}
\affiliation{Donostia International Physics Center (DIPC), E-20018 Donostia/San Sebastian, Spain}
\author{Johannes Feist}
\email{johannes.feist@uam.es}
\affiliation{Departamento de F{\'\i}sica Te{\'o}rica de la Materia Condensada and Condensed Matter Physics Center (IFIMAC), Universidad Aut\'onoma de Madrid, E-28049 Madrid, Spain}

\begin{abstract}
The second law of photochemistry states that in most cases, no more than one
molecule is activated for an excited-state reaction for each photon absorbed by
a collection of molecules. In this work, we demonstrate that it is possible to
trigger a many-molecule reaction using only one photon by strongly coupling the
molecular ensemble to a confined light mode. The collective nature of the
resulting hybrid states of the system (the so-called polaritons) leads to the
formation of a polaritonic ``supermolecule'' involving the degrees of freedom of
all molecules, opening a reaction path on which all involved molecules undergo a
chemical transformation. We theoretically investigate the system conditions for
this effect to take place and be enhanced.
\end{abstract}

\maketitle

Photochemical reactions underlie many essential biological functions,
such as vision or photosynthesis. In this context, the second law of
photochemistry (also known as Stark-Einstein law) states that ``one quantum of
light is absorbed per molecule of absorbing and reacting
substance''~\cite{Rohatgi-Mukherjee2013}. This means that the quantum yield
$\phi=\frac{N_{\mathrm{prod}}}{N_{\mathrm{phot}}}$ of the reaction, which
describes the percentage of molecules that end up in the desired reaction
product per absorbed photon, has a maximum value of 1. This limit can be
overcome in some specific cases, such as in photochemically induced chain
reactions~\cite{Summers1981,Eves2004}, or in systems that support singlet
fission to create multiple triplet excitons (and thus electron-hole pairs) in
solar cells~\cite{Walker2013,Zirzlmeier2015}. In this work, we demonstrate a
novel and general approach to circumvent the Stark-Einstein law by exploiting
the collective nature of polaritonic (hybrid light-matter) states formed by
bringing a collection of molecules into strong coupling with a confined light
mode. We show that this can allow \emph{many} molecules to undergo a
photochemical reaction after excitation by just a \emph{single} photon.
Polaritonic chemistry, i.e., the potential to manipulate chemical structure and
reactions through the formation of polaritons was experimentally demonstrated in
2012~\cite{Hutchison2012}, and has become the topic of intense experimental and
theoretical research in the past few years~\cite{Wang2014Phase, Galego2015,
Galego2016, Herrera2016, Kowalewski2016, Ebbesen2016, Baieva2017, Flick2017,
Kowalewski2017}. However, existing applications and proposals have been limited
to enhancing or suppressing the rates of single-molecule reactions.  In
contrast, we here show that polaritonic chemistry can open fundamentally new
pathways that allow for reactions that are not present in the uncoupled system,
and thus possesses the potential to unlock a new class of collective reactions.

\begin{figure}
  \includegraphics[width=\linewidth]{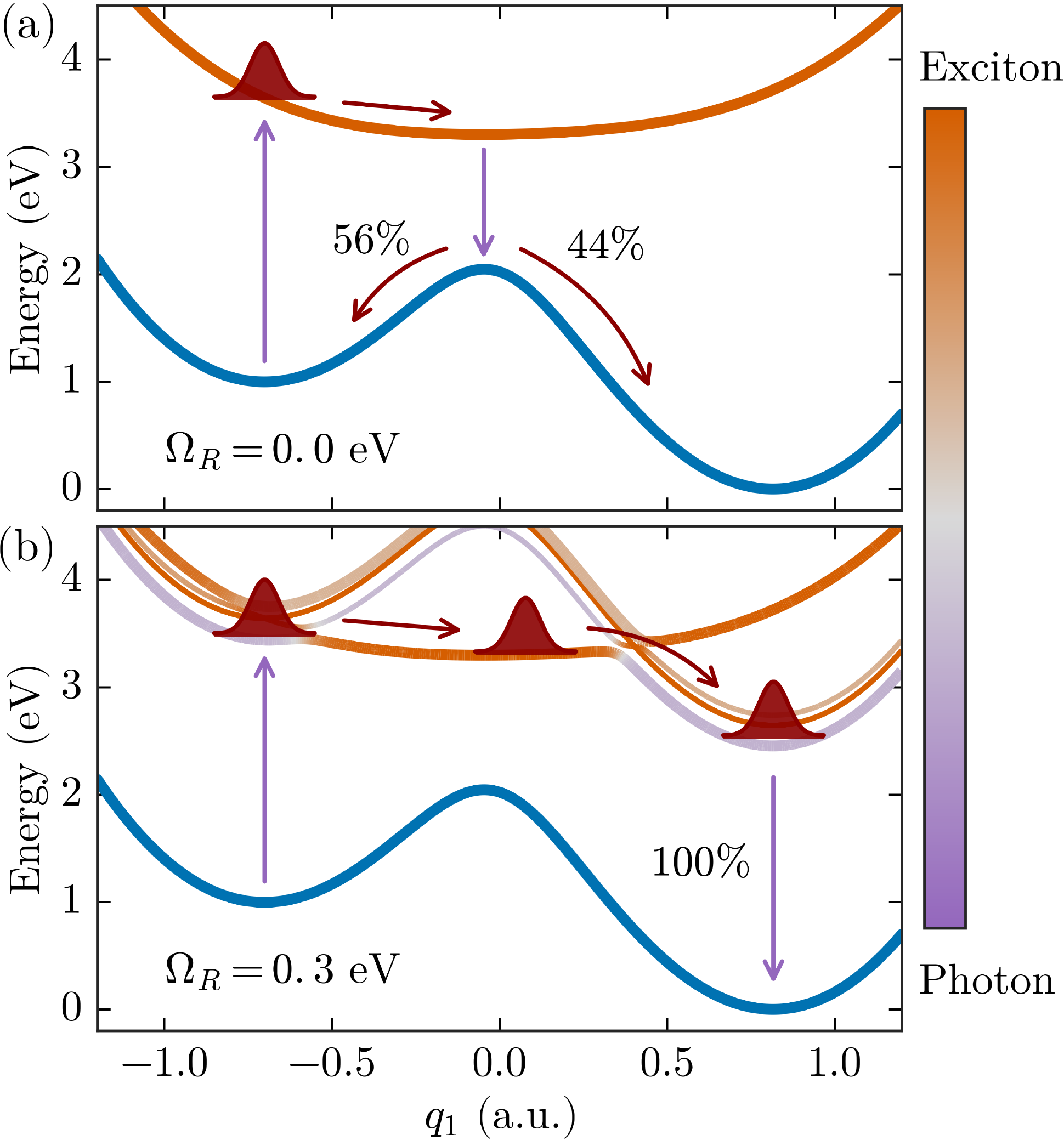}
  \caption{Potential energy surfaces of a system with $N=5$ molecules and one
    light mode under motion of one molecule (a) without light-matter coupling,
    and (b) in the strong coupling regime for $\Omega_R=0.3$~eV at the initial
    position ($q_1=q_{\mathrm{ms}}\approx-0.7$~a.u.). The photon energy is
    $\omega_c= 2.55$~eV. The color scale represents the cavity mode fraction of
    the excited states, going from pure photon (purple) to pure exciton
    (orange).}
  \label{fig:fig1}
\end{figure}

The mechanism we introduce relies on the delocalized character of the hybrid
light-matter excitations (polaritons) formed under strong coupling, which
conceptually leads to the formation of a single ``supermolecule'' involving all
the molecules as well as the trapped photon. The photochemical reaction induced
in this ``supermolecule'' can then lead to most or even all of its molecular
constituents undergoing a structural change, corresponding to an effective
quantum yield significantly larger than $1$, although it can obviously not lead
to a violation of energy conservation. We thus investigate a class of reactions
that release energy, i.e., where the initial starting state after absorption of
a photon has higher energy than the final state, in which all involved molecules
have undergone a reaction. We focus on a class of model molecules with a
structure as proposed for use in solar energy
storage~\cite{Kucharski2011,Cacciarini2015,Gurke2017}, described within a
simplified model treating a single reaction coordinate, as shown in
\autoref{fig:fig1}a. In our model molecule, the electronic ground state contains
two local minima: a stable ground-state configuration (at
$q=q_{\mathrm{s}}\approx 0.8$~a.u.) and a metastable configuration (at
$q=q_{\mathrm{ms}}\approx-0.7$~a.u.) which contains a stored energy of about
$1$~eV. The activation barrier for thermal relaxation from the metastable
configuration to the global minimum has a height of more than $1$~eV, leading to
a lifetime on the order of days or even years for the metastable
configuration~\cite{Eyring1935}, and thus making it interesting for solar energy
storage. In addition, the molecule possesses an excited state with a relatively
flat potential minimum close to the ground transition state. As indicated in
\autoref{fig:fig1}a, vibrational relaxation on the excited-state surface and
subsequent radiative decay gives a roughly equal quantum yield for reaching
either the stable or the metastable configuration in the electronic ground
state. As expected in a conventional photochemical reaction, the quantum yields
in the bare molecule add up to one (indeed, the Stark-Einstein law can be
reformulated as ``the sum of quantum yields must be unity'').

We now assume that a collection of these molecules is placed inside a photonic
structure supporting a single confined light mode. This could be physically
realized using a variety of approaches, such as dielectric microcavities,
photonic crystal cavities, or plasmonic nanocavities~\cite{Spillane2002,
Akahane2003,Daskalakis2014,Lodahl2015,Zengin2015,Chikkaraddy2016,Ramezani2017}.
We analyze the system using the methods developed in our previous work
\cite{Galego2015,Galego2016}, which describe the collective structure of the
coupled system of many-molecules and the photon mode through the use of
polaritonic potential energy surfaces (PES) encompassing the nuclear degrees of
freedom of \emph{all} molecules. This picture immediately allows to take into
account and understand the collective motion induced on the molecules through
their mutual coupling to the photonic mode. For the specific case of five
molecules, we show the coupled PES in \autoref{fig:fig1}b. Here, we take a cut
of the five-dimensional PES where only the first molecule ($q_1$) is allowed to
move, while all others are fixed to the equilibrium position of the metastable
ground-state configuration ($q_i=q_{\mathrm{ms}}$ for $i=2,\ldots,5$).

Already for the motion of just a single
molecule, our results show that the quantum yield for the energy-releasing
back-reaction can be significantly enhanced under strong coupling. The
lowest-energy excited PES (see \autoref{fig:fig1}b) is formed by hybridization
of the uncoupled excited-state surfaces of the molecules with the surface
representing a photon in the cavity and the molecule in the ground state (a copy
of the ground-state surface shifted upwards by the photon energy of the confined
light mode). The photon energy ($\omega_c=2.55$~eV) is close to resonant with
the electronic excitation energy at the metastable configuration
($q=q_\mathrm{ms}$), while most other molecular configurations (and
specifically, the stable configuration $q=q_\mathrm{s}$) are out of resonance
with the cavity. This implies that the nature of the lowest excited-state PES
changes depending on the molecular position $q$, corresponding to a polariton in
some cases, and corresponding to a bare molecular state in others (as indicated
by the color scale in \autoref{fig:fig1}). In the polaritonic states, each
molecule is in its electronic ground state most of the time (since the
excitation is distributed over all the molecules and the photonic mode), such
that the polaritonic parts of the lowest excited-state PES inherit their shape
mostly from the ground-state PES~\cite{Galego2016}. This leads to the formation
of a new minimum in the lowest excited PES at the same position as the fully
relaxed ground-state minimum $q_{\mathrm{s}}$. However, in contrast to the
ground-state PES, there is no significant activation barrier to vibrational
relaxation within the lowest excited-state PES. Instead, a ``bridge'' is formed,
with a shape closely resembling that of the bare-molecule excited-state PES due
to the fact that the cavity is far-detuned from the excited state in that area
of configuration space (around $q_1\approx0$). In the absence of barriers, a
molecular system will quickly relax to the lowest-energy vibrational state on
the lowest excited-state PES, according to Kasha's rule~\cite{Kasha1950}. The
vibrational relaxation in the lowest excited hybrid light-matter PES will thus
lead to localization of the nuclear wave packet close to the ground-state
minimum $q_{\mathrm{s}}$. If radiative decay happens at this point, this would
give a quantum yield of essentially unity for the back-reaction from the
metastable to the stable configuration. While this already presents a large
cavity-induced change of the photochemical properties of such molecules, we next
show that the collective nature of the polaritons can result in an even more
dramatic qualitative change that allows the system to keep releasing energy,
with all molecules relaxing from the metastable to the stable configuration one
after the other.

\begin{figure*}
  \includegraphics[width=\linewidth]{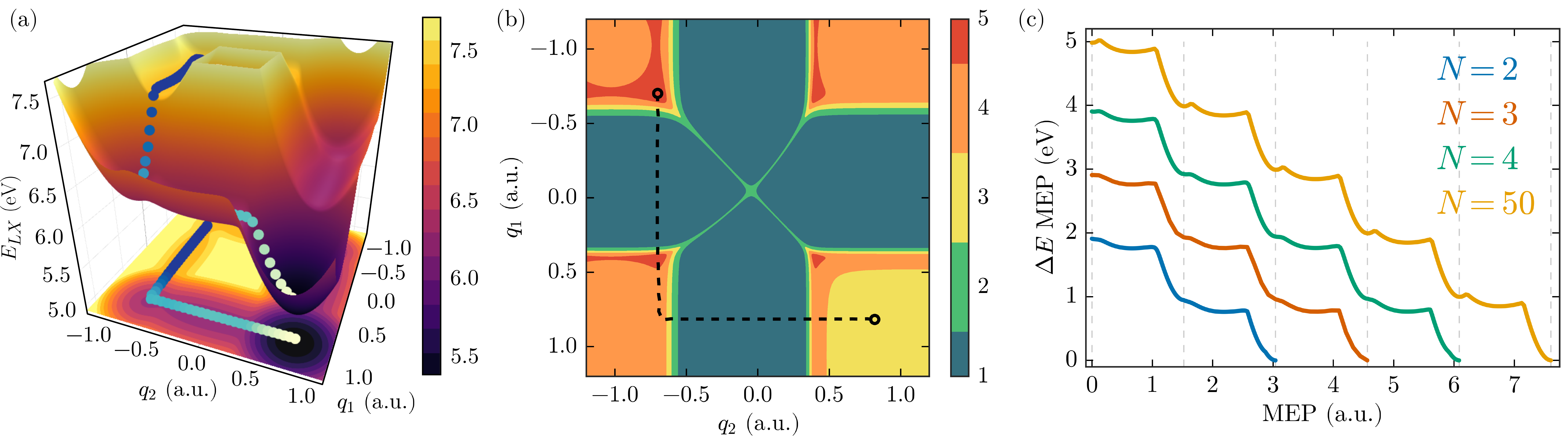}
  \caption{(a) Lowest-energy excited state PES for 2 moving molecules in a
    5-molecule ensemble. The minimum energy path (blue to white dots) connects
    the initial excited region with the final configuration of the two
    molecules. (b) Participation ratio map of the lowest-energy excited state,
    indicating over how many molecules the state is delocalized. The MEP is
    indicated by a dashed black line. (c) Energy profile along the minimum
    energy paths for collections of 2, 3, 4 and 50 molecules. For $N=50$, only
    the first five steps are shown explicitly. Thin dashed lines indicate the
    approximate location along the path where one molecule stops moving and the
    next one starts.}
  \label{fig:fig2}
\end{figure*}

To understand this, we have to take into
account that the polaritonic PESs formed under strong coupling encompass the
nuclear degrees of freedom of \emph{all} involved
molecules~\cite{Galego2015,Galego2016}. This collective nature can in particular
also allow nuclear motion on \emph{different} molecules to become coupled, and
in the current case creates a reaction path along which the system can release
the energy stored in all molecules, while staying on a single adiabatic PES
reached by single-photon absorption in the initial state.  This is demonstrated
for motion of two of the involved molecules in \autoref{fig:fig2}a, which shows
a two-dimensional cut of the PES of the lowest-energy excited state (with all
other molecules again frozen in the metastable position $q=q_{\mathrm{ms}}$). We
calculate the minimum energy path (MEP) connecting the initial configuration
$q_1=q_2=q_{\mathrm{ms}}$ to the location where the first two molecules have
released their stored energy ($q_1=q_2=q_{\mathrm{s}}$) using the nudged elastic
band method~\cite{Henkelman2000}. Along this path, indicated as a series of
points in \autoref{fig:fig2}a, there are no significant reaction barriers, such
that vibrational relaxation after absorption of a single photon indeed can lead
to deactivation of both molecules. The MEP also demonstrates that, to a good
approximation, the reaction proceeds in steps, with the molecules moving one
after the other (i.e., in the first leg, only $q_1$ changes, while in the second
leg, only $q_2$ changes). We note here that due to the indistinguishability of
the molecules, there is an equivalent path where the order of motion of the
molecules is reversed. In order to gain additional insight into the properties
of the polariton states that enable this step-wise many-molecule reaction
triggered by a single photon, we further analyze the lowest excited PES by
showing its molecular participation ratio in \autoref{fig:fig2}b. Here, the
molecular participation ratio is defined as~\cite{Kramer1993}
\begin{equation}\label{eq:PR}
   P_{\alpha}(\vec q) = \frac{\sum_i |\langle e_i | \Psi_{\alpha}(\vec q) \rangle |^2}
                             {\sum_i |\langle e_i | \Psi_{\alpha}(\vec q) \rangle |^4} \,,
\end{equation}
where $|e_i\rangle$ denotes the excited state of molecule $i$, and the sums are
over all molecules. The participation ratio gives an estimate of the number of
molecular states that possess a significant weight in a given state
$|\Psi_{\alpha}\rangle$, with possible values ranging from $P_{\alpha} = 1$ to
$P_{\alpha} = N$ (for $N$ molecules). Analyzing it for the lowest-energy excited
state PES (see \autoref{fig:fig2}b) demonstrates that the surface at the
starting point corresponds to a collective polariton, with the excitation
equally distributed over all molecules. Along the MEP, the excitation collapses
onto a single molecule (the one that is moving), demonstrated by the
participation ratio decreasing to $1$ for $-0.5~\mathrm{a.u.}\lesssim
q_1\lesssim 0.4~\mathrm{a.u.}$. As the molecule moves, it again enters into
resonance with the cavity (and the other molecules) and the state changes
character to a fully delocalized polariton with $P_{LX}=N$ (at $q_1\approx
0.45$~a.u.). However, as the first molecule keeps moving, it falls out of
resonance again and effectively ``drops out'' of the polaritonic state, leaving
the excitation in a polaritonic state distributed over the photonic mode and the
remaining $N-1$ molecules ($P_{LX}=4$), which then forms the starting point for
the second molecule to undergo the reaction. Following the MEP along the second
leg (where $q_1\approx q_{\mathrm{s}}$ and $q_2$ moves from $q_{\mathrm{ms}}$ to
$q_{\mathrm{s}}$), the same process repeats, but now involving one less
molecule.

We now demonstrate that the same process can keep repeating for many molecules.
To this end, we calculate the MEP for varying numbers of molecules from $N=2$ to
$N=50$, with a collective Rabi splitting of $\Omega_R=0.3$~eV in the initial
molecular configuration ($q_i=q_{\mathrm{ms}}$ for all $i$) for all cases. As
shown in \autoref{fig:fig2}c, the energy profile along the MEP is structurally
similar for any number of molecules. The main change is that for larger values
of $N$, the collective protection effect demonstrated in \cite{Galego2016} makes
the PES resemble the shape of the uncoupled PES more strongly, leading to a less
smooth MEP with slightly higher barriers, comparable to the average thermal
kinetic energy at room temperature. In addition, the significant change of
collective state when passing the barrier (with the excitation collapsing from
all molecules onto a single one along the ``bridge'') leads to narrow avoided
crossings in the adiabatic picture. In a diabatic picture formed by the purely
polaritonic PES and the single-molecule excited PES, their coupling can be shown
to scale with $N^{-1/2}$~\cite{Galego2016} (for a fixed collective Rabi
frequency), such that the transition rate to enter into and leave from the
``bridge'' part of the lowest excited-state PES scales with $N^{-1}$. However,
there are many equivalent paths in each step (as any of the remaining molecules
could undergo the reaction), removing the overall scaling with $N$ for the
probability of passing over an arbitrary one of the multiple equivalent barriers
and implies that using standard transition state theory provides a good
approximation for the average time taken to pass the involved barriers. This
gives $\tau \approx \frac{h}{k_BT} \exp\left(\frac{\Delta E}{k_BT}\right)
\lesssim 1$~ps at room temperature~\cite{Eyring1935}, even without taking into
account that the wavepacket arrives at the second, higher, barrier with
significant kinetic energy. This highlights the importance of the lifetime of
the hybrid light-matter states to determine the feasibility of triggering
multiple reactions with a single photon. In most current experiments, polariton
lifetimes (which are an average of the lifetimes of their constituents) are very
short, on the order of tens of femtoseconds, due to the use of short-lived
photonic modes such as localized surface plasmons or the modes of Fabry-Perot
cavities formed from thin metal mirrors. In contrast, the lifetime of the
molecular excitations can be limited by their spontaneous radiative decay, which
is on the order of nanoseconds for typical organic molecules. Consequently, if
long-lived photonic modes as available in low-loss dielectric structures such as
photonic crystals or microtoroidal cavities are used instead, there is no
fundamental reason preventing polariton lifetimes that approach nanoseconds.
This would thus give enough time for thousands of molecules to undergo a
reaction before the excitation is lost due to radiative decay.

In conclusion, we have demonstrated that under strong coupling, a single photon
could be used to trigger a photochemical reaction in \emph{many} molecules. This
corresponds to an effective quantum yield (number of reactant molecules per
absorbed photon) of the reaction that is significantly larger than one, and thus
provides a possible pathway to break the second (or Stark-Einstein) law of
photochemistry. The basic physical effect responsible for this surprising
feature is the delocalized nature of the polaritonic states obtained under
collective strong coupling, which require a treatment of the whole collection of
molecules as a single polaritonic ``supermolecule''. For the specific model
studied here, this strategy could resolve one of the main problems of solar
energy storage: How to efficiently retrieve the stored energy from molecules
that are designed for the opposite purpose, i.e., for storing energy very
efficiently under normal conditions~\cite{Cacciarini2015,Gurke2017}. By
reversibly bringing the system into strong coupling (e.g., through a moving
mirror that brings the cavity into and out of resonance), one could thus trigger
the release the stored energy through absorption of a single ambient photon.

\begin{acknowledgments}
We thank G.~Groenhof for helpful discussions. This work has been funded
by the European Research Council under grant agreements ERC-2011-AdG-290981 and
ERC-2016-STG-714870, by the European Union Seventh Framework Programme under
grant agreement FP7-PEOPLE-2013-CIG-618229, and the Spanish MINECO under
contract MAT2014-53432-C5-5-R and the ``María de Maeztu'' program for Units of
Excellence in R\&D (MDM-2014-0377).
\end{acknowledgments}

\appendix*
\section{Appendix: Theoretical model}
We here describe the molecular model in more detail. The adiabatic PESs of the
single bare molecule are both constructed independently from two coupled
harmonic potentials as follows:
\begin{equation}
V_i(q) = \frac{1}{2} \left(v_i(q) + w_i(q) - \sqrt{4 h_i^2 + \left[v_i(q) - w_i(q)\right]^2}\right),
\end{equation}
where $i\in\{\mathrm{gs, es}\}$ indicates either the ground state or excited
state PES. The terms $v_i(q)$ and $w_i(q)$, coupled through $h_i$, are harmonic
(i.e., quadratic) potentials. For the case of $V_\mathrm{gs}(q)$ the harmonic
potentials are exactly centered at $q_\mathrm{s}$ and $q_\mathrm{ms}$,
respectively, while the ones for $V_\mathrm{es}(q)$ are slightly offset with
respect to these configurations. Although there is a relatively large number of
free parameters that control the molecular details, we have checked the
robustness of our results against reasonable variations.

When introducing the quantized light mode to the system, its coupling to the
molecules is accounted for through an additional term in the Hamiltonian, given
by the scalar product of the molecular dipole moment $\hat{\vec{\mu}}(q)$ and
the EM single-photon electric field strength $\vec{E}_\mathrm{1ph}$. For
simplicity, we use a dipole moment that is constant along the reaction
coordinate and assume perfect alignment between the molecular dipoles and the
electric field direction. We have checked that these assumptions do not
significantly affect the results presented above. The electronic and photonic
part of the full $N$-molecule system Hamiltonian (which determines the hybrid
light-matter PES on which nuclear motion takes place) is then given by:
\begin{equation}
\hat{H}_\mathrm{SC}(\vec{q}) = \omega_c\hat{a}^\dagger \hat{a} +
   \sum_i \left(\hat{V}(q_i) + \hat{\vec{\mu}}(q_i) \cdot \vec{E}_\mathrm{1ph} (\hat{a}^\dagger + \hat{a}) \right),
\end{equation}
where $\hat{V}$ is the diagonal electronic potential operator with the
previously calculated PESs $V_\mathrm{gs}(q)$ and $V_\mathrm{es}(q)$, and
$\hat{a}^\dagger$ and $\hat{a}$ are the bosonic creation and annihilation
operators associated with the confined light mode of energy $\omega_c$.
Diagonalization of $\hat{H}_\mathrm{SC}$ within the single-excitation subspace
yields the polaritonic PES. The collective coupling strength can then be
parametrized through the collective Rabi frequency $\Omega_\mathrm{R} =
2\sqrt{N}\,\vec{\mu} \cdot \vec{E}_\mathrm{1ph}$.

To understand the approximate classical trajectory that defines the reaction
coordinate of the full ``supermolecule'' system, we calculate the minimum energy
path between the initial and final configurations using the nudged elastic band
method. For the initial position, we assume that all molecules are at
$q=q_\mathrm{ms}$, corresponding to short-pulse excitation from the ground state
in the metastable configuration, according to the Franck-Condon principle. For
the final position, we use $q_i=q_\mathrm{s}$ for all $i$, i.e. the position
where all the molecules are in the stable configuration (corresponding to the
global minimum of the PES). Due to the indistinguishability of our molecules,
any of the available molecules can undergo the reaction in each step, and
there are $N!$ equivalent paths from the initial to the final position. In the
results presented above, we thus show only one of these equivalent paths (the
one in which the order of reactions corresponds to the numbering of the
molecules).

\bibliography{references}

\begin{thebibliography}{29}%
\makeatletter
\providecommand \@ifxundefined [1]{%
 \@ifx{#1\undefined}
}%
\providecommand \@ifnum [1]{%
 \ifnum #1\expandafter \@firstoftwo
 \else \expandafter \@secondoftwo
 \fi
}%
\providecommand \@ifx [1]{%
 \ifx #1\expandafter \@firstoftwo
 \else \expandafter \@secondoftwo
 \fi
}%
\providecommand \natexlab [1]{#1}%
\providecommand \enquote  [1]{``#1''}%
\providecommand \bibnamefont  [1]{#1}%
\providecommand \bibfnamefont [1]{#1}%
\providecommand \citenamefont [1]{#1}%
\providecommand \href@noop [0]{\@secondoftwo}%
\providecommand \href [0]{\begingroup \@sanitize@url \@href}%
\providecommand \@href[1]{\@@startlink{#1}\@@href}%
\providecommand \@@href[1]{\endgroup#1\@@endlink}%
\providecommand \@sanitize@url [0]{\catcode `\\12\catcode `\$12\catcode
  `\&12\catcode `\#12\catcode `\^12\catcode `\_12\catcode `\%12\relax}%
\providecommand \@@startlink[1]{}%
\providecommand \@@endlink[0]{}%
\providecommand \url  [0]{\begingroup\@sanitize@url \@url }%
\providecommand \@url [1]{\endgroup\@href {#1}{\urlprefix }}%
\providecommand \urlprefix  [0]{URL }%
\providecommand \Eprint [0]{\href }%
\providecommand \doibase [0]{http://dx.doi.org/}%
\providecommand \selectlanguage [0]{\@gobble}%
\providecommand \bibinfo  [0]{\@secondoftwo}%
\providecommand \bibfield  [0]{\@secondoftwo}%
\providecommand \translation [1]{[#1]}%
\providecommand \BibitemOpen [0]{}%
\providecommand \bibitemStop [0]{}%
\providecommand \bibitemNoStop [0]{.\EOS\space}%
\providecommand \EOS [0]{\spacefactor3000\relax}%
\providecommand \BibitemShut  [1]{\csname bibitem#1\endcsname}%
\let\auto@bib@innerbib\@empty
\bibitem [{\citenamefont {Rohatgi-Mukherjee}(2013)}]{Rohatgi-Mukherjee2013}%
  \BibitemOpen
  \bibfield  {author} {\bibinfo {author} {\bibfnamefont {K.~K.}\ \bibnamefont
  {Rohatgi-Mukherjee}},\ }\href@noop {} {\emph {\bibinfo {title} {{Fundamentals
  of photochemistry}}}}\ (\bibinfo  {publisher} {New Age International},\
  \bibinfo {year} {2013})\ p.\ \bibinfo {pages} {370}\BibitemShut {NoStop}%
\bibitem [{\citenamefont {Summers}\ \emph {et~al.}(1981)\citenamefont
  {Summers}, \citenamefont {Luong},\ and\ \citenamefont
  {Wrighton}}]{Summers1981}%
  \BibitemOpen
  \bibfield  {author} {\bibinfo {author} {\bibfnamefont {David~P.}\
  \bibnamefont {Summers}}, \bibinfo {author} {\bibfnamefont {John~C.}\
  \bibnamefont {Luong}}, \ and\ \bibinfo {author} {\bibfnamefont {Mark~S.}\
  \bibnamefont {Wrighton}},\ }\bibfield  {title} {\enquote {\bibinfo {title}
  {{A New Mechanism for Photosubstitution of Organometallic Complexes.
  Generation of Substitutionally Labile Oxidation States by Excited-State
  Electron Transfer in the Presence of Ligands}},}\ }\href {\doibase
  10.1021/ja00407a050} {\bibfield  {journal} {\bibinfo  {journal} {J. Am. Chem.
  Soc.}\ }\textbf {\bibinfo {volume} {103}},\ \bibinfo {pages} {5238} (\bibinfo
  {year} {1981})}\BibitemShut {NoStop}%
\bibitem [{\citenamefont {Eves}\ \emph {et~al.}(2004)\citenamefont {Eves},
  \citenamefont {Sun}, \citenamefont {Lopinski},\ and\ \citenamefont
  {Zuilhof}}]{Eves2004}%
  \BibitemOpen
  \bibfield  {author} {\bibinfo {author} {\bibfnamefont {Brian~J.}\
  \bibnamefont {Eves}}, \bibinfo {author} {\bibfnamefont {Qiao-Yu}\
  \bibnamefont {Sun}}, \bibinfo {author} {\bibfnamefont {Gregory~P.}\
  \bibnamefont {Lopinski}}, \ and\ \bibinfo {author} {\bibfnamefont {Han}\
  \bibnamefont {Zuilhof}},\ }\bibfield  {title} {\enquote {\bibinfo {title}
  {{Photochemical Attachment of Organic Monolayers onto H-Terminated Si(111):
  Radical Chain Propagation Observed via STM Studies}},}\ }\href {\doibase
  10.1021/ja045777x} {\bibfield  {journal} {\bibinfo  {journal} {J. Am. Chem.
  Soc.}\ }\textbf {\bibinfo {volume} {126}},\ \bibinfo {pages} {14318}
  (\bibinfo {year} {2004})}\BibitemShut {NoStop}%
\bibitem [{\citenamefont {Walker}\ \emph {et~al.}(2013)\citenamefont {Walker},
  \citenamefont {Musser}, \citenamefont {Beljonne},\ and\ \citenamefont
  {Friend}}]{Walker2013}%
  \BibitemOpen
  \bibfield  {author} {\bibinfo {author} {\bibfnamefont {Brian~J}\ \bibnamefont
  {Walker}}, \bibinfo {author} {\bibfnamefont {Andrew~J}\ \bibnamefont
  {Musser}}, \bibinfo {author} {\bibfnamefont {David}\ \bibnamefont
  {Beljonne}}, \ and\ \bibinfo {author} {\bibfnamefont {Richard~H}\
  \bibnamefont {Friend}},\ }\bibfield  {title} {\enquote {\bibinfo {title}
  {{Singlet exciton fission in solution}},}\ }\href {\doibase
  10.1038/nchem.1801} {\bibfield  {journal} {\bibinfo  {journal} {Nat. Chem.}\
  }\textbf {\bibinfo {volume} {5}},\ \bibinfo {pages} {1019} (\bibinfo {year}
  {2013})}\BibitemShut {NoStop}%
\bibitem [{\citenamefont {Zirzlmeier}\ \emph {et~al.}(2015)\citenamefont
  {Zirzlmeier}, \citenamefont {Lehnherr}, \citenamefont {Coto}, \citenamefont
  {Chernick}, \citenamefont {Casillas}, \citenamefont {Basel}, \citenamefont
  {Thoss}, \citenamefont {Tykwinski},\ and\ \citenamefont
  {Guldi}}]{Zirzlmeier2015}%
  \BibitemOpen
  \bibfield  {author} {\bibinfo {author} {\bibfnamefont {Johannes}\
  \bibnamefont {Zirzlmeier}}, \bibinfo {author} {\bibfnamefont {Dan}\
  \bibnamefont {Lehnherr}}, \bibinfo {author} {\bibfnamefont {Pedro~B}\
  \bibnamefont {Coto}}, \bibinfo {author} {\bibfnamefont {Erin~T}\ \bibnamefont
  {Chernick}}, \bibinfo {author} {\bibfnamefont {Rub{\'{e}}n}\ \bibnamefont
  {Casillas}}, \bibinfo {author} {\bibfnamefont {Bettina~S}\ \bibnamefont
  {Basel}}, \bibinfo {author} {\bibfnamefont {Michael}\ \bibnamefont {Thoss}},
  \bibinfo {author} {\bibfnamefont {Rik~R}\ \bibnamefont {Tykwinski}}, \ and\
  \bibinfo {author} {\bibfnamefont {Dirk~M}\ \bibnamefont {Guldi}},\ }\bibfield
   {title} {\enquote {\bibinfo {title} {{Singlet fission in pentacene
  dimers}},}\ }\href {\doibase 10.1073/pnas.1422436112} {\bibfield  {journal}
  {\bibinfo  {journal} {Proc. Natl. Acad. Sci.}\ }\textbf {\bibinfo {volume}
  {112}},\ \bibinfo {pages} {5325} (\bibinfo {year} {2015})}\BibitemShut
  {NoStop}%
\bibitem [{\citenamefont {Hutchison}\ \emph {et~al.}(2012)\citenamefont
  {Hutchison}, \citenamefont {Schwartz}, \citenamefont {Genet}, \citenamefont
  {Devaux},\ and\ \citenamefont {Ebbesen}}]{Hutchison2012}%
  \BibitemOpen
  \bibfield  {author} {\bibinfo {author} {\bibfnamefont {James~A.}\
  \bibnamefont {Hutchison}}, \bibinfo {author} {\bibfnamefont {Tal}\
  \bibnamefont {Schwartz}}, \bibinfo {author} {\bibfnamefont {Cyriaque}\
  \bibnamefont {Genet}}, \bibinfo {author} {\bibfnamefont {Elo{\"{i}}se}\
  \bibnamefont {Devaux}}, \ and\ \bibinfo {author} {\bibfnamefont {Thomas~W.}\
  \bibnamefont {Ebbesen}},\ }\bibfield  {title} {\enquote {\bibinfo {title}
  {{Modifying Chemical Landscapes by Coupling to Vacuum Fields}},}\ }\href
  {\doibase 10.1002/ange.201107033} {\bibfield  {journal} {\bibinfo  {journal}
  {Angew. Chemie}\ }\textbf {\bibinfo {volume} {124}},\ \bibinfo {pages} {1624}
  (\bibinfo {year} {2012})}\BibitemShut {NoStop}%
\bibitem [{\citenamefont {Wang}\ \emph {et~al.}(2014)\citenamefont {Wang},
  \citenamefont {Mika}, \citenamefont {Hutchison}, \citenamefont {Genet},
  \citenamefont {Jouaiti}, \citenamefont {Hosseini},\ and\ \citenamefont
  {Ebbesen}}]{Wang2014Phase}%
  \BibitemOpen
  \bibfield  {author} {\bibinfo {author} {\bibfnamefont {Shaojun}\ \bibnamefont
  {Wang}}, \bibinfo {author} {\bibfnamefont {Arkadiusz}\ \bibnamefont {Mika}},
  \bibinfo {author} {\bibfnamefont {James~A.}\ \bibnamefont {Hutchison}},
  \bibinfo {author} {\bibfnamefont {Cyriaque}\ \bibnamefont {Genet}}, \bibinfo
  {author} {\bibfnamefont {Abdelaziz}\ \bibnamefont {Jouaiti}}, \bibinfo
  {author} {\bibfnamefont {Mir~Wais}\ \bibnamefont {Hosseini}}, \ and\ \bibinfo
  {author} {\bibfnamefont {Thomas~W.}\ \bibnamefont {Ebbesen}},\ }\bibfield
  {title} {\enquote {\bibinfo {title} {{Phase transition of a perovskite
  strongly coupled to the vacuum field}},}\ }\href {\doibase
  10.1039/c4nr01971g} {\bibfield  {journal} {\bibinfo  {journal} {Nanoscale}\
  }\textbf {\bibinfo {volume} {6}},\ \bibinfo {pages} {7243} (\bibinfo {year}
  {2014})}\BibitemShut {NoStop}%
\bibitem [{\citenamefont {Galego}\ \emph {et~al.}(2015)\citenamefont {Galego},
  \citenamefont {Garcia-Vidal},\ and\ \citenamefont {Feist}}]{Galego2015}%
  \BibitemOpen
  \bibfield  {author} {\bibinfo {author} {\bibfnamefont {Javier}\ \bibnamefont
  {Galego}}, \bibinfo {author} {\bibfnamefont {Francisco~J.}\ \bibnamefont
  {Garcia-Vidal}}, \ and\ \bibinfo {author} {\bibfnamefont {Johannes}\
  \bibnamefont {Feist}},\ }\bibfield  {title} {\enquote {\bibinfo {title}
  {{Cavity-Induced Modifications of Molecular Structure in the Strong-Coupling
  Regime}},}\ }\href {\doibase 10.1103/PhysRevX.5.041022} {\bibfield  {journal}
  {\bibinfo  {journal} {Phys. Rev. X}\ }\textbf {\bibinfo {volume} {5}},\
  \bibinfo {pages} {041022} (\bibinfo {year} {2015})}\BibitemShut {NoStop}%
\bibitem [{\citenamefont {Galego}\ \emph {et~al.}(2016)\citenamefont {Galego},
  \citenamefont {Garcia-Vidal},\ and\ \citenamefont {Feist}}]{Galego2016}%
  \BibitemOpen
  \bibfield  {author} {\bibinfo {author} {\bibfnamefont {Javier}\ \bibnamefont
  {Galego}}, \bibinfo {author} {\bibfnamefont {Francisco~J.}\ \bibnamefont
  {Garcia-Vidal}}, \ and\ \bibinfo {author} {\bibfnamefont {Johannes}\
  \bibnamefont {Feist}},\ }\bibfield  {title} {\enquote {\bibinfo {title}
  {{Suppressing photochemical reactions with quantized light fields}},}\ }\href
  {\doibase 10.1038/ncomms13841} {\bibfield  {journal} {\bibinfo  {journal}
  {Nat. Commun.}\ }\textbf {\bibinfo {volume} {7}},\ \bibinfo {pages} {13841}
  (\bibinfo {year} {2016})}\BibitemShut {NoStop}%
\bibitem [{\citenamefont {Herrera}\ and\ \citenamefont
  {Spano}(2016)}]{Herrera2016}%
  \BibitemOpen
  \bibfield  {author} {\bibinfo {author} {\bibfnamefont {Felipe}\ \bibnamefont
  {Herrera}}\ and\ \bibinfo {author} {\bibfnamefont {Frank~C.}\ \bibnamefont
  {Spano}},\ }\bibfield  {title} {\enquote {\bibinfo {title}
  {{Cavity-Controlled Chemistry in Molecular Ensembles}},}\ }\href {\doibase
  10.1103/PhysRevLett.116.238301} {\bibfield  {journal} {\bibinfo  {journal}
  {Phys. Rev. Lett.}\ }\textbf {\bibinfo {volume} {116}},\ \bibinfo {pages}
  {238301} (\bibinfo {year} {2016})}\BibitemShut {NoStop}%
\bibitem [{\citenamefont {Kowalewski}\ \emph {et~al.}(2016)\citenamefont
  {Kowalewski}, \citenamefont {Bennett},\ and\ \citenamefont
  {Mukamel}}]{Kowalewski2016}%
  \BibitemOpen
  \bibfield  {author} {\bibinfo {author} {\bibfnamefont {Markus}\ \bibnamefont
  {Kowalewski}}, \bibinfo {author} {\bibfnamefont {Kochise}\ \bibnamefont
  {Bennett}}, \ and\ \bibinfo {author} {\bibfnamefont {Shaul}\ \bibnamefont
  {Mukamel}},\ }\bibfield  {title} {\enquote {\bibinfo {title} {{Non-adiabatic
  dynamics of molecules in optical cavities}},}\ }\href {\doibase
  10.1063/1.4941053} {\bibfield  {journal} {\bibinfo  {journal} {J. Chem.
  Phys.}\ }\textbf {\bibinfo {volume} {144}},\ \bibinfo {pages} {054309}
  (\bibinfo {year} {2016})}\BibitemShut {NoStop}%
\bibitem [{\citenamefont {Ebbesen}(2016)}]{Ebbesen2016}%
  \BibitemOpen
  \bibfield  {author} {\bibinfo {author} {\bibfnamefont {Thomas~W.}\
  \bibnamefont {Ebbesen}},\ }\bibfield  {title} {\enquote {\bibinfo {title}
  {{Hybrid Light–Matter States in a Molecular and Material Science
  Perspective}},}\ }\href {\doibase 10.1021/acs.accounts.6b00295} {\bibfield
  {journal} {\bibinfo  {journal} {Acc. Chem. Res.}\ }\textbf {\bibinfo {volume}
  {49}},\ \bibinfo {pages} {2403} (\bibinfo {year} {2016})}\BibitemShut
  {NoStop}%
\bibitem [{\citenamefont {Baieva}\ \emph {et~al.}(2017)\citenamefont {Baieva},
  \citenamefont {Hakamaa}, \citenamefont {Groenhof}, \citenamefont
  {Heikkil{\"{a}}},\ and\ \citenamefont {Toppari}}]{Baieva2017}%
  \BibitemOpen
  \bibfield  {author} {\bibinfo {author} {\bibfnamefont {Svitlana}\
  \bibnamefont {Baieva}}, \bibinfo {author} {\bibfnamefont {Ossi}\ \bibnamefont
  {Hakamaa}}, \bibinfo {author} {\bibfnamefont {Gerrit}\ \bibnamefont
  {Groenhof}}, \bibinfo {author} {\bibfnamefont {Tero~T.}\ \bibnamefont
  {Heikkil{\"{a}}}}, \ and\ \bibinfo {author} {\bibfnamefont {J.~Jussi}\
  \bibnamefont {Toppari}},\ }\bibfield  {title} {\enquote {\bibinfo {title}
  {{Dynamics of Strongly Coupled Modes between Surface Plasmon Polaritons and
  Photoactive Molecules: The Effect of the Stokes Shift}},}\ }\href {\doibase
  10.1021/acsphotonics.6b00482} {\bibfield  {journal} {\bibinfo  {journal} {ACS
  Photonics}\ }\textbf {\bibinfo {volume} {4}},\ \bibinfo {pages} {28}
  (\bibinfo {year} {2017})}\BibitemShut {NoStop}%
\bibitem [{\citenamefont {Flick}\ \emph {et~al.}(2017)\citenamefont {Flick},
  \citenamefont {Ruggenthaler}, \citenamefont {Appel},\ and\ \citenamefont
  {Rubio}}]{Flick2017}%
  \BibitemOpen
  \bibfield  {author} {\bibinfo {author} {\bibfnamefont {Johannes}\
  \bibnamefont {Flick}}, \bibinfo {author} {\bibfnamefont {Michael}\
  \bibnamefont {Ruggenthaler}}, \bibinfo {author} {\bibfnamefont {Heiko}\
  \bibnamefont {Appel}}, \ and\ \bibinfo {author} {\bibfnamefont {Angel}\
  \bibnamefont {Rubio}},\ }\bibfield  {title} {\enquote {\bibinfo {title}
  {{Atoms and molecules in cavities, from weak to strong coupling in
  quantum-electrodynamics (QED) chemistry}},}\ }\href {\doibase
  10.1073/pnas.1615509114} {\bibfield  {journal} {\bibinfo  {journal} {Proc.
  Natl. Acad. Sci.}\ }\textbf {\bibinfo {volume} {114}},\ \bibinfo {pages}
  {3026} (\bibinfo {year} {2017})}\BibitemShut {NoStop}%
\bibitem [{\citenamefont {Kowalewski}\ and\ \citenamefont
  {Mukamel}(2017)}]{Kowalewski2017}%
  \BibitemOpen
  \bibfield  {author} {\bibinfo {author} {\bibfnamefont {Markus}\ \bibnamefont
  {Kowalewski}}\ and\ \bibinfo {author} {\bibfnamefont {Shaul}\ \bibnamefont
  {Mukamel}},\ }\bibfield  {title} {\enquote {\bibinfo {title} {{Manipulating
  molecules with quantum light}},}\ }\href {\doibase 10.1073/pnas.1702160114}
  {\bibfield  {journal} {\bibinfo  {journal} {Proc. Natl. Acad. Sci.}\ }\textbf
  {\bibinfo {volume} {114}},\ \bibinfo {pages} {3278} (\bibinfo {year}
  {2017})}\BibitemShut {NoStop}%
\bibitem [{\citenamefont {Kucharski}\ \emph {et~al.}(2011)\citenamefont
  {Kucharski}, \citenamefont {Tian}, \citenamefont {Akbulatov},\ and\
  \citenamefont {Boulatov}}]{Kucharski2011}%
  \BibitemOpen
  \bibfield  {author} {\bibinfo {author} {\bibfnamefont {Timothy~J.}\
  \bibnamefont {Kucharski}}, \bibinfo {author} {\bibfnamefont {Yancong}\
  \bibnamefont {Tian}}, \bibinfo {author} {\bibfnamefont {Sergey}\ \bibnamefont
  {Akbulatov}}, \ and\ \bibinfo {author} {\bibfnamefont {Roman}\ \bibnamefont
  {Boulatov}},\ }\bibfield  {title} {\enquote {\bibinfo {title} {{Chemical
  solutions for the closed-cycle storage of solar energy}},}\ }\href {\doibase
  10.1039/c1ee01861b} {\bibfield  {journal} {\bibinfo  {journal} {Energy
  Environ. Sci.}\ }\textbf {\bibinfo {volume} {4}},\ \bibinfo {pages} {4449}
  (\bibinfo {year} {2011})}\BibitemShut {NoStop}%
\bibitem [{\citenamefont {Cacciarini}\ \emph {et~al.}(2015)\citenamefont
  {Cacciarini}, \citenamefont {Skov}, \citenamefont {Jevric}, \citenamefont
  {Hansen}, \citenamefont {Elm}, \citenamefont {Kjaergaard}, \citenamefont
  {Mikkelsen},\ and\ \citenamefont {{Br{\o}ndsted Nielsen}}}]{Cacciarini2015}%
  \BibitemOpen
  \bibfield  {author} {\bibinfo {author} {\bibfnamefont {Martina}\ \bibnamefont
  {Cacciarini}}, \bibinfo {author} {\bibfnamefont {Anders~B.}\ \bibnamefont
  {Skov}}, \bibinfo {author} {\bibfnamefont {Martyn}\ \bibnamefont {Jevric}},
  \bibinfo {author} {\bibfnamefont {Anne~S.}\ \bibnamefont {Hansen}}, \bibinfo
  {author} {\bibfnamefont {Jonas}\ \bibnamefont {Elm}}, \bibinfo {author}
  {\bibfnamefont {Henrik~G.}\ \bibnamefont {Kjaergaard}}, \bibinfo {author}
  {\bibfnamefont {Kurt~V.}\ \bibnamefont {Mikkelsen}}, \ and\ \bibinfo {author}
  {\bibfnamefont {Mogens}\ \bibnamefont {{Br{\o}ndsted Nielsen}}},\ }\bibfield
  {title} {\enquote {\bibinfo {title} {{Towards Solar Energy Storage in the
  Photochromic Dihydroazulene-Vinylheptafulvene System}},}\ }\href {\doibase
  10.1002/chem.201500100} {\bibfield  {journal} {\bibinfo  {journal} {Chem.
  Eur. J}\ }\textbf {\bibinfo {volume} {21}},\ \bibinfo {pages} {7454}
  (\bibinfo {year} {2015})}\BibitemShut {NoStop}%
\bibitem [{\citenamefont {Gurke}\ \emph {et~al.}(2017)\citenamefont {Gurke},
  \citenamefont {Quick}, \citenamefont {Ernsting},\ and\ \citenamefont
  {Hecht}}]{Gurke2017}%
  \BibitemOpen
  \bibfield  {author} {\bibinfo {author} {\bibfnamefont {J.}~\bibnamefont
  {Gurke}}, \bibinfo {author} {\bibfnamefont {M.}~\bibnamefont {Quick}},
  \bibinfo {author} {\bibfnamefont {N.~P.}\ \bibnamefont {Ernsting}}, \ and\
  \bibinfo {author} {\bibfnamefont {S.}~\bibnamefont {Hecht}},\ }\bibfield
  {title} {\enquote {\bibinfo {title} {{Acid-catalysed thermal cycloreversion
  of a diarylethene: a potential way for triggered release of stored light
  energy?}}}\ }\href {\doibase 10.1039/C6CC10182H} {\bibfield  {journal}
  {\bibinfo  {journal} {Chem. Commun.}\ }\textbf {\bibinfo {volume} {53}},\
  \bibinfo {pages} {2150} (\bibinfo {year} {2017})}\BibitemShut {NoStop}%
\bibitem [{\citenamefont {Eyring}(1935)}]{Eyring1935}%
  \BibitemOpen
  \bibfield  {author} {\bibinfo {author} {\bibfnamefont {Henry}\ \bibnamefont
  {Eyring}},\ }\bibfield  {title} {\enquote {\bibinfo {title} {{The Activated
  Complex in Chemical Reactions}},}\ }\href {\doibase 10.1063/1.1749604}
  {\bibfield  {journal} {\bibinfo  {journal} {J. Chem. Phys.}\ }\textbf
  {\bibinfo {volume} {3}},\ \bibinfo {pages} {107} (\bibinfo {year}
  {1935})}\BibitemShut {NoStop}%
\bibitem [{\citenamefont {Spillane}\ \emph {et~al.}(2002)\citenamefont
  {Spillane}, \citenamefont {Kippenberg},\ and\ \citenamefont
  {Vahala}}]{Spillane2002}%
  \BibitemOpen
  \bibfield  {author} {\bibinfo {author} {\bibfnamefont {S.~M.}\ \bibnamefont
  {Spillane}}, \bibinfo {author} {\bibfnamefont {T.~J.}\ \bibnamefont
  {Kippenberg}}, \ and\ \bibinfo {author} {\bibfnamefont {K.~J.}\ \bibnamefont
  {Vahala}},\ }\bibfield  {title} {\enquote {\bibinfo {title}
  {{Ultralow-threshold Raman laser using a spherical dielectric
  microcavity}},}\ }\href {\doibase 10.1038/415621a} {\bibfield  {journal}
  {\bibinfo  {journal} {Nature}\ }\textbf {\bibinfo {volume} {415}},\ \bibinfo
  {pages} {621} (\bibinfo {year} {2002})}\BibitemShut {NoStop}%
\bibitem [{\citenamefont {Akahane}\ \emph {et~al.}(2003)\citenamefont
  {Akahane}, \citenamefont {Asano}, \citenamefont {Song},\ and\ \citenamefont
  {Noda}}]{Akahane2003}%
  \BibitemOpen
  \bibfield  {author} {\bibinfo {author} {\bibfnamefont {Yoshihiro}\
  \bibnamefont {Akahane}}, \bibinfo {author} {\bibfnamefont {Takashi}\
  \bibnamefont {Asano}}, \bibinfo {author} {\bibfnamefont {Bong-Shik}\
  \bibnamefont {Song}}, \ and\ \bibinfo {author} {\bibfnamefont {Susumu}\
  \bibnamefont {Noda}},\ }\bibfield  {title} {\enquote {\bibinfo {title}
  {{High-Q photonic nanocavity in a two-dimensional photonic crystal}},}\
  }\href {\doibase 10.1038/nature02063} {\bibfield  {journal} {\bibinfo
  {journal} {Nature}\ }\textbf {\bibinfo {volume} {425}},\ \bibinfo {pages}
  {944} (\bibinfo {year} {2003})}\BibitemShut {NoStop}%
\bibitem [{\citenamefont {Daskalakis}\ \emph {et~al.}(2014)\citenamefont
  {Daskalakis}, \citenamefont {Maier}, \citenamefont {Murray},\ and\
  \citenamefont {K{\'{e}}na-Cohen}}]{Daskalakis2014}%
  \BibitemOpen
  \bibfield  {author} {\bibinfo {author} {\bibfnamefont {K.~S.}\ \bibnamefont
  {Daskalakis}}, \bibinfo {author} {\bibfnamefont {S.~A.}\ \bibnamefont
  {Maier}}, \bibinfo {author} {\bibfnamefont {R.}~\bibnamefont {Murray}}, \
  and\ \bibinfo {author} {\bibfnamefont {S.}~\bibnamefont {K{\'{e}}na-Cohen}},\
  }\bibfield  {title} {\enquote {\bibinfo {title} {{Nonlinear interactions in
  an organic polariton condensate}},}\ }\href {\doibase 10.1038/nmat3874}
  {\bibfield  {journal} {\bibinfo  {journal} {Nat. Mater.}\ }\textbf {\bibinfo
  {volume} {13}},\ \bibinfo {pages} {271} (\bibinfo {year} {2014})}\BibitemShut
  {NoStop}%
\bibitem [{\citenamefont {Lodahl}\ \emph {et~al.}(2015)\citenamefont {Lodahl},
  \citenamefont {Mahmoodian},\ and\ \citenamefont {Stobbe}}]{Lodahl2015}%
  \BibitemOpen
  \bibfield  {author} {\bibinfo {author} {\bibfnamefont {Peter}\ \bibnamefont
  {Lodahl}}, \bibinfo {author} {\bibfnamefont {Sahand}\ \bibnamefont
  {Mahmoodian}}, \ and\ \bibinfo {author} {\bibfnamefont {S{\o}ren}\
  \bibnamefont {Stobbe}},\ }\bibfield  {title} {\enquote {\bibinfo {title}
  {{Interfacing single photons and single quantum dots with photonic
  nanostructures}},}\ }\href {\doibase 10.1103/RevModPhys.87.347} {\bibfield
  {journal} {\bibinfo  {journal} {Rev. Mod. Phys.}\ }\textbf {\bibinfo {volume}
  {87}},\ \bibinfo {pages} {347} (\bibinfo {year} {2015})}\BibitemShut
  {NoStop}%
\bibitem [{\citenamefont {Zengin}\ \emph {et~al.}(2015)\citenamefont {Zengin},
  \citenamefont {Wers{\"{a}}ll}, \citenamefont {Nilsson}, \citenamefont
  {Antosiewicz}, \citenamefont {K{\"{a}}ll},\ and\ \citenamefont
  {Shegai}}]{Zengin2015}%
  \BibitemOpen
  \bibfield  {author} {\bibinfo {author} {\bibfnamefont {G{\"{u}}lis}\
  \bibnamefont {Zengin}}, \bibinfo {author} {\bibfnamefont {Martin}\
  \bibnamefont {Wers{\"{a}}ll}}, \bibinfo {author} {\bibfnamefont {Sara}\
  \bibnamefont {Nilsson}}, \bibinfo {author} {\bibfnamefont {Tomasz~J.}\
  \bibnamefont {Antosiewicz}}, \bibinfo {author} {\bibfnamefont {Mikael}\
  \bibnamefont {K{\"{a}}ll}}, \ and\ \bibinfo {author} {\bibfnamefont {Timur}\
  \bibnamefont {Shegai}},\ }\bibfield  {title} {\enquote {\bibinfo {title}
  {{Realizing Strong Light-Matter Interactions between Single-Nanoparticle
  Plasmons and Molecular Excitons at Ambient Conditions}},}\ }\href {\doibase
  10.1103/PhysRevLett.114.157401} {\bibfield  {journal} {\bibinfo  {journal}
  {Phys. Rev. Lett.}\ }\textbf {\bibinfo {volume} {114}},\ \bibinfo {pages}
  {157401} (\bibinfo {year} {2015})}\BibitemShut {NoStop}%
\bibitem [{\citenamefont {Chikkaraddy}\ \emph {et~al.}(2016)\citenamefont
  {Chikkaraddy}, \citenamefont {de~Nijs}, \citenamefont {Benz}, \citenamefont
  {Barrow}, \citenamefont {Scherman}, \citenamefont {Rosta}, \citenamefont
  {Demetriadou}, \citenamefont {Fox}, \citenamefont {Hess},\ and\ \citenamefont
  {Baumberg}}]{Chikkaraddy2016}%
  \BibitemOpen
  \bibfield  {author} {\bibinfo {author} {\bibfnamefont {Rohit}\ \bibnamefont
  {Chikkaraddy}}, \bibinfo {author} {\bibfnamefont {Bart}\ \bibnamefont
  {de~Nijs}}, \bibinfo {author} {\bibfnamefont {Felix}\ \bibnamefont {Benz}},
  \bibinfo {author} {\bibfnamefont {Steven~J.}\ \bibnamefont {Barrow}},
  \bibinfo {author} {\bibfnamefont {Oren~A.}\ \bibnamefont {Scherman}},
  \bibinfo {author} {\bibfnamefont {Edina}\ \bibnamefont {Rosta}}, \bibinfo
  {author} {\bibfnamefont {Angela}\ \bibnamefont {Demetriadou}}, \bibinfo
  {author} {\bibfnamefont {Peter}\ \bibnamefont {Fox}}, \bibinfo {author}
  {\bibfnamefont {Ortwin}\ \bibnamefont {Hess}}, \ and\ \bibinfo {author}
  {\bibfnamefont {Jeremy~J.}\ \bibnamefont {Baumberg}},\ }\bibfield  {title}
  {\enquote {\bibinfo {title} {{Single-molecule strong coupling at room
  temperature in plasmonic nanocavities}},}\ }\href {\doibase
  10.1038/nature17974} {\bibfield  {journal} {\bibinfo  {journal} {Nature}\
  }\textbf {\bibinfo {volume} {535}},\ \bibinfo {pages} {127} (\bibinfo {year}
  {2016})}\BibitemShut {NoStop}%
\bibitem [{\citenamefont {Ramezani}\ \emph {et~al.}(2017)\citenamefont
  {Ramezani}, \citenamefont {Halpin}, \citenamefont
  {Fern{\'{a}}ndez-Dom{\'{i}}nguez}, \citenamefont {Feist}, \citenamefont
  {Rodriguez}, \citenamefont {Garcia-Vidal},\ and\ \citenamefont {{G{\'{o}}mez
  Rivas}}}]{Ramezani2017}%
  \BibitemOpen
  \bibfield  {author} {\bibinfo {author} {\bibfnamefont {Mohammad}\
  \bibnamefont {Ramezani}}, \bibinfo {author} {\bibfnamefont {Alexei}\
  \bibnamefont {Halpin}}, \bibinfo {author} {\bibfnamefont {Antonio~I.}\
  \bibnamefont {Fern{\'{a}}ndez-Dom{\'{i}}nguez}}, \bibinfo {author}
  {\bibfnamefont {Johannes}\ \bibnamefont {Feist}}, \bibinfo {author}
  {\bibfnamefont {Said Rahimzadeh-Kalaleh}\ \bibnamefont {Rodriguez}}, \bibinfo
  {author} {\bibfnamefont {Francisco~J.}\ \bibnamefont {Garcia-Vidal}}, \ and\
  \bibinfo {author} {\bibfnamefont {Jaime}\ \bibnamefont {{G{\'{o}}mez
  Rivas}}},\ }\bibfield  {title} {\enquote {\bibinfo {title}
  {{Plasmon-exciton-polariton lasing}},}\ }\href {\doibase
  10.1364/OPTICA.4.000031} {\bibfield  {journal} {\bibinfo  {journal} {Optica}\
  }\textbf {\bibinfo {volume} {4}},\ \bibinfo {pages} {31} (\bibinfo {year}
  {2017})}\BibitemShut {NoStop}%
\bibitem [{\citenamefont {Kasha}(1950)}]{Kasha1950}%
  \BibitemOpen
  \bibfield  {author} {\bibinfo {author} {\bibfnamefont {Michael}\ \bibnamefont
  {Kasha}},\ }\bibfield  {title} {\enquote {\bibinfo {title} {{Characterization
  of electronic transitions in complex molecules}},}\ }\href {\doibase
  10.1039/df9500900014} {\bibfield  {journal} {\bibinfo  {journal} {Discuss.
  Faraday Soc.}\ }\textbf {\bibinfo {volume} {9}},\ \bibinfo {pages} {14}
  (\bibinfo {year} {1950})}\BibitemShut {NoStop}%
\bibitem [{\citenamefont {Henkelman}\ \emph {et~al.}(2000)\citenamefont
  {Henkelman}, \citenamefont {Uberuaga},\ and\ \citenamefont
  {J{\'{o}}nsson}}]{Henkelman2000}%
  \BibitemOpen
  \bibfield  {author} {\bibinfo {author} {\bibfnamefont {Graeme}\ \bibnamefont
  {Henkelman}}, \bibinfo {author} {\bibfnamefont {Blas~P.}\ \bibnamefont
  {Uberuaga}}, \ and\ \bibinfo {author} {\bibfnamefont {Hannes}\ \bibnamefont
  {J{\'{o}}nsson}},\ }\bibfield  {title} {\enquote {\bibinfo {title} {{A
  climbing image nudged elastic band method for finding saddle points and
  minimum energy paths}},}\ }\href {\doibase 10.1063/1.1329672} {\bibfield
  {journal} {\bibinfo  {journal} {J. Chem. Phys.}\ }\textbf {\bibinfo {volume}
  {113}},\ \bibinfo {pages} {9901} (\bibinfo {year} {2000})}\BibitemShut
  {NoStop}%
\bibitem [{\citenamefont {Kramer}\ and\ \citenamefont
  {MacKinnon}(1993)}]{Kramer1993}%
  \BibitemOpen
  \bibfield  {author} {\bibinfo {author} {\bibfnamefont {Bernhard}\
  \bibnamefont {Kramer}}\ and\ \bibinfo {author} {\bibfnamefont {Angus}\
  \bibnamefont {MacKinnon}},\ }\bibfield  {title} {\enquote {\bibinfo {title}
  {{Localization: theory and experiment}},}\ }\href {\doibase
  10.1088/0034-4885/56/12/001} {\bibfield  {journal} {\bibinfo  {journal} {Rep.
  Prog. Phys.}\ }\textbf {\bibinfo {volume} {56}},\ \bibinfo {pages} {1469}
  (\bibinfo {year} {1993})}\BibitemShut {NoStop}%
\end{thebibliography}%

\end{document}